\documentclass{IEEEtran4PSCC}
\usepackage{cite}
\usepackage{amssymb,amsfonts}
\usepackage{textcomp}
\usepackage[T1]{fontenc}
\usepackage{beramono}
\usepackage{listings}
\usepackage[usenames,dvipsnames]{xcolor}

\usepackage{float}
\usepackage{algorithm}
\usepackage{algorithmicx}
\usepackage{caption}
\usepackage{subcaption}
\usepackage{enumitem}

\ifCLASSINFOpdf
   \usepackage[pdftex]{graphicx}
\else
   \usepackage[dvips]{graphicx}
\fi
\usepackage[cmex10]{amsmath}
\makeatletter
\let\old@ps@headings\ps@headings
\let\old@ps@IEEEtitlepagestyle\ps@IEEEtitlepagestyle
\def\psccfooter#1{%
    \def\ps@headings{%
        \old@ps@headings%
        \def\@oddfoot{\strut\hfill#1\hfill\strut}%
        \def\@evenfoot{\strut\hfill#1\hfill\strut}%
    }%
    \def\ps@IEEEtitlepagestyle{%
        \old@ps@IEEEtitlepagestyle%
        \def\@oddfoot{\strut\hfill#1\hfill\strut}%
        \def\@evenfoot{\strut\hfill#1\hfill\strut}%
    }%
    \ps@headings%
}
\makeatother

\begin{document}
%
\title{Long Duration Battery Sizing, Siting, and Operation Under Wildfire Risk Using Progressive Hedging}

 \author{\IEEEauthorblockN{Ryan Piansky,\IEEEauthorrefmark{1}\textsuperscript{\textsection}
 Georgia Stinchfield,\IEEEauthorrefmark{2}\textsuperscript{\textsection}
 Alyssa Kody,\IEEEauthorrefmark{3} 
 Daniel K. Molzahn,\IEEEauthorrefmark{1} and
 Jean-Paul Watson\IEEEauthorrefmark{4}
 }
 \IEEEauthorblockA{\IEEEauthorrefmark{1} School of Electrical and Computer Engineering,
 Georgia Institute of Technology\\
 Atlanta, Georgia USA, \{rpiansky3, molzahn\}@gatech.edu}
  \IEEEauthorblockA{\IEEEauthorrefmark{2}Department of Chemical Engineering,
 Carnegie Mellon University\\
 Pittsburgh, Pennsylvania USA, gstinchf@andrew.cmu.edu}
 \IEEEauthorblockA{\IEEEauthorrefmark{3} Energy Systems and Infrastructure Analysis, 
 Argonne National Laboratory\\
 Lemont, Illinois USA, akody@anl.gov}
 \IEEEauthorblockA{\IEEEauthorrefmark{4}
 Lawrence Livermore National Laboratory\\
 Livermore, California USA, watson61@llnl.gov}
 }

\maketitle
\begingroup\renewcommand\thefootnote{\textsection}
\footnotetext{G. Stinchfield and R. Piansky are co-first authors.}
\endgroup

\begin{abstract}
Battery sizing and siting problems are computationally challenging due to the need to make long-term planning decisions that are cognizant of short-term operational decisions. This paper considers sizing, siting, and operating batteries in a power grid to maximize their benefits, including price arbitrage and load shed mitigation, during both normal operations and periods with high wildfire ignition risk. We formulate a multi-scenario optimization problem for long duration battery storage while considering the possibility of load shedding during Public Safety Power Shutoff (PSPS) events that de-energize lines to mitigate severe wildfire ignition risk. To enable a computationally scalable solution of this problem with many scenarios of wildfire risk and power injection variability, we develop a customized temporal decomposition method based on a progressive hedging framework. Extending traditional progressive hedging techniques, we consider coupling in both placement variables across all scenarios and state-of-charge variables at temporal boundaries. This enforces consistency across scenarios while enabling parallel computations despite both spatial and temporal coupling. The proposed decomposition facilitates efficient and scalable modeling of a full year of hourly operational decisions to inform the sizing and siting of batteries. With this decomposition, we model a year of hourly operational decisions to inform optimal battery placement for a 240-bus WECC model in under 70 minutes of wall-clock time.

\end{abstract}


\section{Introduction}

Wildfires pose a growing threat due to accelerating climate change \cite{martinuzzi2019future}. While power system infrastructure is only responsible for a small fraction of all wildfire ignitions, these often lead to more severe and expansive fires compared to other ignition sources \cite{keeley2019twenty}. 
In parts of the Western United States during periods of elevated wildfire threat, system operators proactively de-energize power lines with the goal of reducing the risk of wildfire ignition. These de-energizations are referred to as  ``Public Safety Power Shutoff'' (PSPS) events~\cite{pge2021PSPS}. When a PSPS event is invoked, system operators identify power lines to de-energize that pose a high ignition risk. These lines are selected using information about the lines (e.g., condition, age, routing), environmental factors (e.g., humidity, wind, temperature), wildfire spread models, etc. \cite{sotolongo2020california}.

Although PSPS events can mitigate acute ignition risk, they also can result in load shedding as de-energized portions of the grid cannot serve all load demands.
Furthermore, power outages can have significant negative economic and societal impacts~\cite{wong2022support}.
To aid in reducing the impact or extent of PSPS events, utilities are investing in various types of infrastructure such as undergrounded lines~\cite{pge2023undergrounding}, microgrids~\cite{taylor2023managing}, and covered conductors~\cite{pge2023hardening} as well as vegetation management~\cite{pge2023hardening}. 
Utility-scale battery energy storage systems~\cite{yang2018,marnell2019transmission} are another investment that can help mitigate the impact of PSPS events on customers during wildfire season~\cite{singer2023batteries, murray2022caiso}.
However, unlike some of these other investment options, utility-scale batteries can also be beneficial during periods without wildfire ignition threats to improve system operations and renewable integration \cite{santos2022influence}.
This leads to the following question: \textit{How do we optimally size, site, and operate utility-scale batteries on a transmission network considering both high wildfire-risk periods during which PSPS events are enacted and also normal operations?}

During normal operations, we may aim to locate batteries in congested areas of the network or near large concentrations of renewables~\cite{fiorini2017sizing}.
During high wildfire-risk conditions, the topology of the network will change due to PSPS de-energization events. Therefore, congested areas of the network may shift and some parts of the network may experience load shedding. These seasonal changes make it difficult to identify appropriate sizing and siting decisions without considering long, multi-month time horizons. Simultaneously, to properly evaluate the benefits of battery systems, one must consider decisions at the operational time scale, such as hourly charging and discharging behavior. This paper proposes a method to take into account a \textit{full year of hourly operation decisions} spanning both wildfire and non-wildfire threat time periods.

Finding optimal locations to place utility-scale batteries on a large power grid is a computationally challenging problem~\cite{aguado2017battery, santos2022influence, fiorini2017sizing}; see~\cite{yang2018, marnell2019transmission} for literature reviews. Given the need to consider detailed operational decisions across multiple scenarios, the tractability of this problem scales poorly in both the number of scenarios and grid size.
Prior work has developed algorithms using smaller test systems, shorter time horizons, or larger operational time steps~\cite{aguado2017battery, santos2022influence}. Work has also been done to model battery behavior in conjunction with an AC power flow network representation, but again, on smaller test networks~\cite{fiorini2017sizing}. Stochastic programming techniques have been used for sizing and siting problems to consider variability in renewable forecasts~\cite{chen2019optimal}. Similarly, a decomposition method based on progressive hedging has been used for problems related to wildfire operational problems that optimize line switching decisions~\cite{zhou2023optimal}, but these problems did not consider batteries. The need to model both extreme wildfire conditions with PSPS events and normal operations requires a level of temporal granularity and length that further challenges the optimization of battery sizing and siting. 

In this paper, we propose a method for optimally sizing, siting, and operating long duration utility-scale battery systems considering both PSPS events and normal operations.
To enable computational tractability, we employ a progressive hedging (PH) algorithm, which is a scenario-based decomposition technique originally developed for stochastic programming~\cite{watson2011progressive}. Traditionally, to model uncertainties, PH algorithms consider many scenarios that are all coupled by one set of common variables. In contrast, since our setting instead considers a temporal decomposition which represents a year of hourly operations via week-long scenarios, our algorithm must consider both coupling with a common set of variables and coupling between scenarios at temporal boundaries. Specifically, these scenarios must agree on (1)~the sizing and siting of utility-scale batteries, and (2)~the state-of-charge values at temporal boundaries. This setting thus requires non-trivial modifications to standard progressive hedging algorithms to account for this additional temporal coupling.

For week-long scenarios representing periods during wildfire season, sufficiently high wildfire risks result in line de-energizations.
We model the wildfire risk posed by each line using real wildfire risk data from United States Geological Survey (USGS).
Our approach allows us to model long-duration energy storage while capturing the different wildfire and load conditions at various times over the year.

The optimization model for the placement problem considering multiple scenarios is defined in Section~\ref{sec:modeling}, while our solution approach using a customized progressive hedging algorithm is detailed in Section~\ref{sec:ph-algorithm}. Our experimental methodology is described in Section~\ref{sec:methodology} followed by discussions of specific case studies and results in Section~\ref{sec:results}. We conclude in Section~\ref{sec:conclusion} with a summary of our contributions and a discussion of future work.

\section{Optimization Model}
\label{sec:modeling}

We now describe our long duration battery sizing, siting, and operational optimization model. We set a 100MW per-unit (p.u.) basis.

\subsection{Network, Demand, and Generator Variables}

\noindent We define index sets over a transmission network as follows:
\begin{itemize}[noitemsep]
    \item $\mathcal{N}$, the set of buses
    \item $\mathcal{L}$, the set of transmission lines
    \item $\mathcal{G}$, the set of generators
    \item $\mathcal{T} = \{1, \ldots, T \}$,  the (ordered) set of time periods
\end{itemize}

\noindent For each line $\ell \in \mathcal{L}$, we specify the following parameters:
\begin{itemize}[noitemsep]
    \item $b^\ell$, the line susceptance in p.u.
    \item $\overline{p}^\ell$, the power flow limit in p.u. 
    \item $r^\ell_t$, the unitless wildfire risk associated with line $\ell$ being energized at time $t$
    \item $n^{\ell, \text{fr}}$ and $n^{\ell, \text{to}}$,  \textit{to} and \textit{from} buses, respectively, where positive power flows from the \textit{from} bus to the \textit{to} bus
    \item $\overline{\delta}^\ell$ and $\underline{\delta}^\ell$, upper and lower voltage angle difference limits in radians, respectively
    \item $\ell \in \mathcal{L}^{\text{off}}_t \subseteq \mathcal{L}$, a subset of lines that are de-energized (switched off) during time $t$. Note: the line's energization state is dictated by the wildfire risk on a given day and is thus not constant for all $t \in \mathcal{T}$
\end{itemize}

\noindent For each generator $i \in \mathcal{G}$, we specify the following parameters:
\begin{itemize}[noitemsep]
    \item $\overline{g}^i$ and $\underline{g}^i$, upper and lower generation limits, respectively, in p.u.
    \item $n^i$, bus $n$ at which generator $i$ is located 
    \item $c_j^i$, the $j^{th}$ order cost term for generator $i$ with $j \in \{1, 2, ..., J\}$ for a provided $J^{th}$ degree polynomial cost function
\end{itemize}

\noindent For each bus $n \in \mathcal{N}$, we specify the following parameters:
\begin{itemize}[noitemsep]
    \item $p_{l,t}^n$, power demand at time $t \in \mathcal{T}$ in p.u.
    \item $\mathcal{G}^n$, the set of generators located at bus $n$
    \item $\mathcal{L}^{n, {\text{to}}}$ and $\mathcal{L}^{n, {\text{fr}}}$, the subset of lines $\ell \in \mathcal{L}$ with bus $n$ as the designated \textit{to} bus, and bus $n$ as the designated \textit{from} bus, respectively
\end{itemize}

\noindent We model the operational problem for each time period using a B-$\theta$ DC optimal power flow representation, necessitating the following variables:
\begin{itemize}[noitemsep]
    \item $p_{g,t}^i$, output of generator $i \in \mathcal{G}$ at time $t \in \mathcal{T}$ in p.u.
    \item $g_{slack,t}^n$, slack generation at bus $n \in \mathcal{N}$ at time $t \in \mathcal{T}$ in p.u.
    \item $\theta^n_t$, voltage angle, in radians, at bus $n \in \mathcal{N}$ at time $t \in \mathcal{T}$
    \item $p_{ls,t}^n$, load shedding at bus $n \in N$ at time $t \in \mathcal{T}$ in p.u.
    \item $p^\ell_t$, power flow from bus $n^{\ell, \text{fr}}$ to bus $n^{\ell, \text{to}}$ on line $\ell \in \mathcal{L}$ at time $t \in \mathcal{T}$ in p.u.
\end{itemize}

\subsection{Network Operational Bounds and Constraints}

We use a B-$\theta$ DC OPF with the following constraints

\begin{subequations} \label{eq: dcots}
    \begin{align}
        &\underline{g}^i \leqslant p_{g,t}^i \leqslant \overline{g}^i, & \forall i \in \mathcal{G}, \forall t \in \mathcal{T}, \label{const: gen limits}
        \\
        &0 \leqslant p_{ls,t}^n \leqslant p_{l,t}^n, & \forall n \in \mathcal{N}, \ \forall t \in \mathcal{T}, \label{const: load shed limits}
        \\ 
        &0 \leqslant g_{slack,t}^n \leqslant 10000, & \forall n \in \mathcal{N}, \ \forall t \in \mathcal{T}, \label{const: gen slack limits}
        \\ 
       &\underline{\delta}^\ell  \leqslant \theta^{n^{\ell, \text{fr}}}_t - \theta^{n^{\ell, \text{to}}}_t \leqslant \overline{\delta}^\ell, & \forall \ell \in \mathcal{L}\setminus\mathcal{L}^{\text{off}}, \ \forall t \in \mathcal{T}, \label{const: angle limits}
        \\
        &-b^\ell(\theta^{n^{\ell, \text{fr}}} - \theta^{n^{\ell, \text{to}}})  \leqslant p_t^\ell \leqslant -b^\ell(\theta^{n^{\ell, \text{fr}}} - \theta^{n^{\ell, \text{to}}}), \hspace*{-8em} \nonumber \\
        & & \forall \ell \in \mathcal{L}\setminus\mathcal{L}^{\text{off}}_t, \ \forall t \in \mathcal{T}, \label{const: power flow}
    \end{align}
\end{subequations}

\noindent where \eqref{const: gen limits} bounds the output of each generator $i \in \mathcal{G}$ based on its physical limits, \eqref{const: load shed limits} constrains the allowable load shed at each bus based on the load demanded, \eqref{const: angle limits} bounds the voltage angle difference across each energized line, and \eqref{const: power flow} models the power flow on each energized line $\ell \in \mathcal{L}$ using the B-$\theta$ DC approximation. To help with feasibility in the decomposed model described in Sections~\ref{sec:ph-algorithm} and~\ref{sec:methodology}, we introduce slack generation with bounds described in \eqref{const: gen slack limits}.

Power flow on each line is bi-directional with the thermal limit set as a bound, unless the line is de-energized, in which case the power flow is set to $0$:
\begin{align} 
     p^\ell_t = 0, \qquad \forall \ell \in \mathcal{L}^{\text{off}}_t, \ \forall t \in \mathcal{T},\\
    -\overline{p}^\ell \leqslant p^\ell_t \leqslant \overline{p}^\ell, \qquad \forall \ell \in \mathcal{L} \setminus \mathcal{L}^{\text{off}}_t, \ \forall t \in \mathcal{T}. \label{const: powr flow limits}
\end{align}
Note that the line de-energizations are based solely on the time-varying wildfire risks and are not decision variables in the optimization formulation.

\subsection{Grid-Scale Battery Modeling}
\label{sec:battery-model}


We allow batteries to be placed at a subset of buses $\mathcal{N}^{\text{batt}} \subseteq \mathcal{N}$. This allows for flexibility in large systems where a heuristic needs to be used to identify a subset of candidate battery locations to improve tractability. Unless otherwise specified, $\mathcal{N}^{\text{batt}} = \mathcal{N}$. 

Battery parameters are defined as:
\begin{itemize}[noitemsep]
    \item $\overline{E}$ and $\underline{E}$, the upper and lower energy storage limits of the battery, respectively, in p.u.
    \item $E^n_0$, the  initial charge of batteries at bus $n \in \mathcal{N}^{\text{batt}}$ in p.u.
    \item $\overline{p}_c$ and $\underline{p}_c$, the upper and lower charge/discharge rate limits for a single battery, respectively, in p.u. in a single time interval
    \item $e$, the efficiency of charging and $\frac{1}{e}$ the efficiency of discharging
    \item $h$, the hourly carry-over rate to model self-discharge
    \item $X_{max}$, the maximum number of batteries allowed at a single bus
    \item $X_{total}$, the total number of batteries allowed across the entire network
\end{itemize}
For each bus $n \in \mathcal{N}^{\text{batt}}$, we introduce the following variables:
\begin{itemize}[noitemsep]
    \item $x^n \in \mathbb{R}^+$, number of batteries placed at bus $n$
    \item $p_{c,t}^n$, charging rate at bus $n$ at time $t \in \mathcal{T}$ in p.u.
    \item $p_{d,t}^n$, charging rate at bus $n$ at time $t \in \mathcal{T}$ in p.u.
\end{itemize}

\noindent Limits are imposed on the number of batteries introduced to the network and the amount placed at each bus:
\begin{align}\label{const: batt amounts}
    0 \leq x^n &\leq X_{max}, \qquad \forall n \in \mathcal{N}^{\text{batt}},\\
    \sum_{n \in \mathcal{N}^{\text{batt}}} x^n &\leq X_{total}.
\end{align}
Let $E_t^n(\cdot)$ be the total energy, or state-of-charge (SOC), stored in all batteries placed at bus $n \in \mathcal{N}^{\text{batt}}$ at time $t \in \mathcal{T}$. The stored energy changes as the batteries charge and discharge:
\begin{align} \label{const: soc update}
    E_{t+1}^n(x,p_c,p_d) = hE_{t}^n(x,p_c,p_d) + ep_{c,t}^n - \frac{1}{e}p_{d,t}^n
\end{align}
The energy stored in the set of batteries at bus $n \in \mathcal{N}^{\text{batt}}$ must abide by lower and upper storage bounds:
\begin{align} \label{const: charge limits}
    x^n\underline{E} \leqslant E_{t+1}^n(x,p_c,p_d) \leqslant x^n\overline{E}, \qquad \forall n \in \mathcal{N}^{\text{batt}}, \ \forall t \in \mathcal{T}.
\end{align}
Charging and discharging is enforced via lower and upper rate limits:
\begin{align} \label{const: charging rate}
    & x^n\underline{p}_c \leqslant  p_{c,t}^n \leqslant x^n\overline{p}_c, \quad \forall n \in \mathcal{N}^{\text{batt}}, & \forall t \in \mathcal{T},\\
 \label{const: discharging rate}
    &x^n\underline{p}_c \leqslant  p_{d,t}^n \leqslant x^n\overline{p}_c, \quad \forall n \in \mathcal{N}^{\text{batt}}, & \forall t \in \mathcal{T}.
\end{align}

\begin{table}[]
\renewcommand{\arraystretch}{1.3}
\caption{Battery Parameters}\label{table:battery parameters}
\centering
\begin{tabular}{r|c|c|}
\cline{2-3}
                                         & Parameter         & Value     \\ \hline
\multicolumn{1}{|r|}{minimum storage limit}  & $\underline{E}$   & 0 p.u.    \\
\multicolumn{1}{|r|}{maximum storage limit}  & $\overline{E}$    & 1.0 p.u.  \\
\multicolumn{1}{|r|}{minimum charge rate}    & $\underline{p}_c$ & 0 p.u./hour    \\
\multicolumn{1}{|r|}{maximum charge rate}    & $\overline{p}_c$  & 1.0 p.u./hour  \\ 
\multicolumn{1}{|r|}{maximum batteries}    & $X_{max}$  & 4 batteries  \\ 
\multicolumn{1}{|r|}{efficiency}    & $e$  & 95\%  \\
\multicolumn{1}{|r|}{hourly carryover}    & $h$  & 0.999958 \\
\multicolumn{1}{|r|}{total batteries}    & $X_{total}$  & 10 batteries  \\ 
\hline
\end{tabular}
\end{table}

Consistent with the scale of ongoing battery installations~\cite{colthorpe2023us}, Table~\ref{table:battery parameters} summarizes the battery parameters used in our numerical tests.
Batteries are modeled to have a capacity of 1.0 p.u. (100~MWh) in our problem formulation~\cite{sandia}. This is consistent with utility-scale lithium-ion battery installations~\cite{battery_landscape}. 
A charge/discharge efficiency of 95\% (roughly a 90\% round-trip efficiency) is used and an hourly carryover rate is used to model a 0.1\% daily self-discharge loss~\cite{luo2015overview}. Generalizations can be made to introduce other models of losses, such as the formulation in~\cite{nazir2021guaranteeing}, or multiple types of batteries with distinct characteristics. 

With the inclusion of grid-scale batteries, power balance is enforced across all buses at all times:
\begin{multline} \label{const: power balance batts}
    \sum_{\ell \in \mathcal{L}^{n, \text{fr}}} p^\ell_t -\sum_{\ell \in \mathcal{L}^{n, \text{to}}} p^\ell_t = \sum_{i \in \mathcal{G}^n} p_{g,t}^i + g_{slack,t}^n - p_{l,t}^n + p_{ls,t}^n \\[-0.5em] - p_{c,t}^n + p_{d,t}^n,
    \forall n \in \mathcal{N}, \ \forall t \in \mathcal{T}.
\end{multline}

\subsection{Objective Function}


A specified number of batteries are placed in the system with no associated cost. This allows for the optimization to focus on minimizing operational costs over the time horizon,~$\mathcal{T}$. Modeled costs include production for power generation and penalties imposed for unserved load, as follows:
\begin{align}
    C_{gen} &= \sum_{t \in \mathcal{T}} \sum_{i \in \mathcal{G}^n} \sum_{j=0}^{J-1} c_i^j \cdot (p_{g,t}^i)^j, \\
    C_{loadshed} &= \sum_{t \in \mathcal{T}} \sum_{n \in \mathcal{N}} K_{ls} \cdot p_{ls,t}^n. 
    \\
    C_{slack} &= \sum_{t \in \mathcal{T}} \sum_{n \in \mathcal{N}} K_{slack} \cdot g_{slack,t}^n.
\end{align}
\noindent Here, $K_{ls}$ represents the cost per {p.u.}~of demand not served, in this model set to be $\$20,000/\text{p.u.}$~and the cost of slack generation $K_{slack}$ is set to 50$K_{ls}$. The total system opreating cost is then given by:
\begin{align}\label{const: cost total}
    C_{tot} = C_{gen} + C_{loadshed} + C_{slack}.
\end{align}
The optimization problem we formulate for sizing, siting, and operating systems with long duration energy storage is thus:
\begin{equation}
\begin{aligned}
&\min\limits_{p_g, \theta, p^{\ell}, p_{ls}, x, p_b, r_g, r_b} \ \eqref{const: cost total}\\
&\qquad\qquad\text{s.t.} \ \eqref{eq: dcots}\text{\,--\,}\eqref{const: powr flow limits},\; \eqref{const: batt amounts}\text{\,--\,}\eqref{const: power balance batts}. 
\end{aligned}
\label{Cost-Opt}
\end{equation}
Note that this problem takes the form of a linear program (LP).

\subsection{Thresholded Line Switching and Investments}
\label{sec:opf-switching}

Unlike in previous work~\cite{kody2022optimizing}, line de-energization decisions are predetermined by a risk threshold instead of a binary decision variable embedded in the optimization problem. In a similar manner to \cite{kody2022optimizing}, each transmission line in the network is assigned a daily risk of wildfire ignition. In this paper, lines above a set threshold on a given day are de-energized and thus included in $\mathcal{L}^{\text{off}}_t$. Accordingly, the line energization status on a given day is a \emph{parameter} rather than a decision variable such as in, e.g., \cite{taylor2023managing, rhodes2020balancing, kody2022optimizing, kody2022sharing}. Extensions to simultaneously consider long duration battery sizing, siting, and operation along with optimizing line de-energization decisions is a topic of ongoing work.


\section{Decomposition via Progressive Hedging}
\label{sec:ph-algorithm}

Stochastic programs are often employed to enable power grid optimization models to hedge against the prevalence of uncertainties such as load and renewables production, and ultimately make more informed and reliable decisions \cite{roald2023power}. The block-angular structure in deterministic multi-scenario planning problems considering large numbers of time periods (including the model we propose in Section~\ref{sec:modeling}) is identical to that of a stochastic program, and we can thus leverage mathematical machinery from stochastic programming to address our deterministic problem~\eqref{Cost-Opt}. In practice, such problems are typically too large to solve directly without tailored optimization approaches. The Progressive Hedging (PH) algorithm, first proposed in~\cite{rockafellar1991scenarios}, was developed to handle this challenge. PH exploits the structure of stochastic programs by breaking the larger problem into multiple smaller, more computationally tractable sub-problems that can be solved in parallel.

The primary input to a stochastic program is a set of representative scenarios, which we denote $\mathcal{S}{=}\{1,...,S\}$. In our battery optimization model, scenarios correspond to operational time periods. Scenario data in a stochastic program is partitioned into multiple stages, $\mathcal{J}=\{2,...,J\}$. Typically, stage indexes start at $j=2$ because constant, scenario-independent data is given at stage $j=1$. Across all stages and scenarios are realizations of random variables, stored in $\xi$, drawn from the discrete probability space $\Xi$. A stage $j{\in}\mathcal{J}$ represents a point in time across the scenarios where uncertain parameters \emph{up to and including} stage $j$, which corresponds to the vector $\vec{\xi}^{j}(s)$, become known and specified decisions must be made. A scenario $s{\in}\mathcal{S}$ contains a full set of realized random variables $\xi(s)$ for all stages $j{\in}\mathcal{J}$ and is weighted by the probability of that realization occurring, $\mathbb{P}(s)$. In the context of our deterministic battery sizing and siting formulation, we could use $\mathbb{P}(s)$ to weight each time period by its frequency of occurrence if we employed representative weeks to model longer periods. However, in our implementation, we model each week individually and thus weight each period the same. 

Here, we consider a two-stage problem ($\mathcal{J}{=}\{2\}$) in which first-stage decisions (storage sizing and siting) are made before the random variables (wildfire risk and load demand vectors) are realized or observable; all operational decisions are second-stage. 
While we only consider two-stage problems, we note that the modeling approach and solution approach based on PH can easily be extended to the multi-stage case. This could allow for modeling of uncertainties in load and risk profiles within each scenario (week) of the simulation, which is a subject of our future work.

\begin{figure}[t]
    \centering
        \includegraphics[scale=0.33]{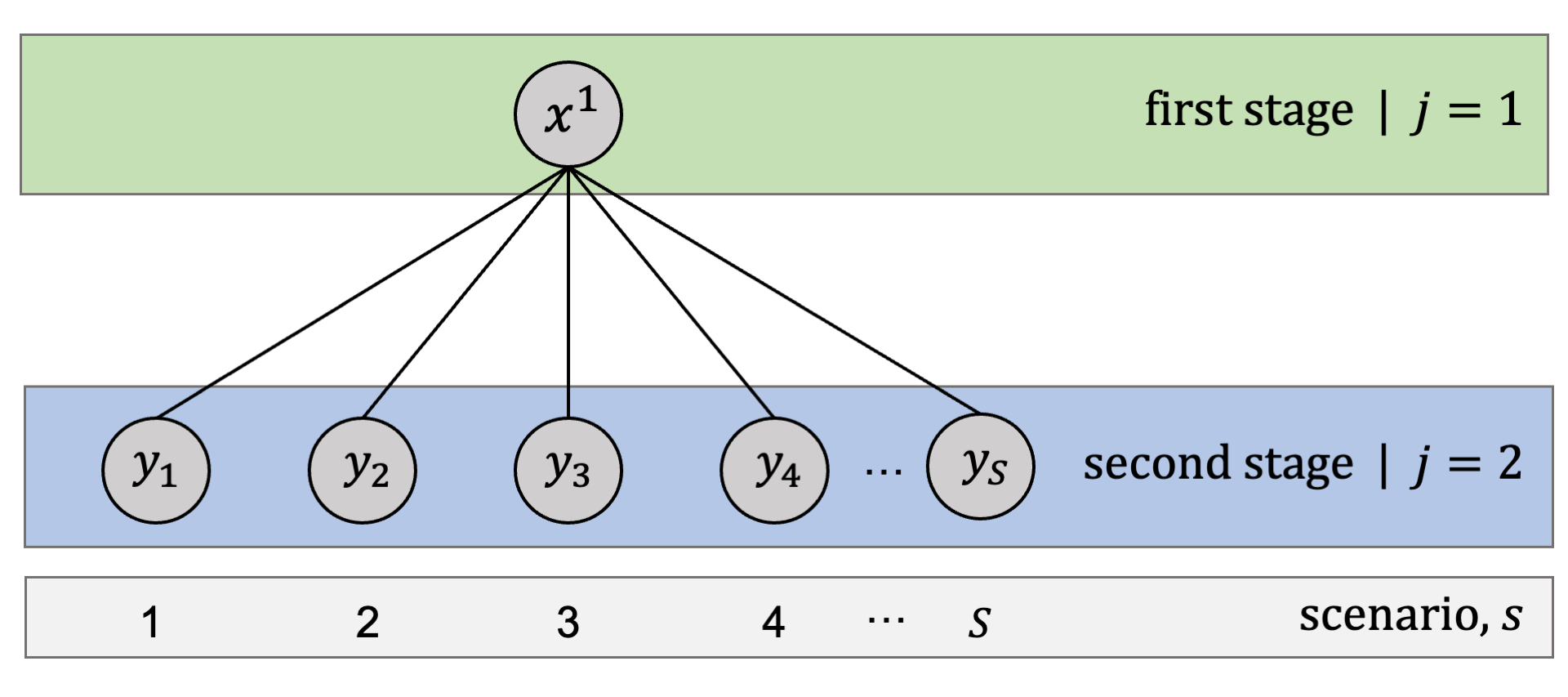}
        \caption{Two-stage stochastic program representation}
        \label{fig:stoprog}
        \vspace{-1em}
\end{figure}

We denote first- and second-stage variables generally as $x{\in}\mathbb{R}^n$ and $y{\in}\mathbb{R}^{S \times m}$, respectively. In a solution to a two-stage stochastic program, first-stage variables are identical across scenarios, while second-stage variables are scenario-dependent. We will denote a scenario-specific set of second-stage variables as $y_s{\in}\mathbb{R}^m$. First-stage variables must be indexed over the stage at which they are made. Let $x^j$ denote the decision vector for stage $j$, $\vec{x}^j$ hold all decision for stages up to and including stage $j$, and $x$ refer to decisions made for all stages in $\mathcal{J}$. This first- and second-stage relationship is shown graphically in Figure~\ref{fig:stoprog}.

The general formulation of a stochastic program is: 
\begin{equation}
\begin{aligned}
    \label{stoprog_obj}
    \min_{x,y_s}
    \sum_{s{\in}\mathcal{S}} \mathbb{P}(s) 
    \bigg{[} 
    f_1\big{(}x^1\big{)} + \sum\limits_{j=2}^{J}  f_j\Big{(}x^j;\vec{x}^{j-1},\vec{\xi}^j(s)\Big{)}
    \bigg{]}\\
    x\in X_{\xi},\; \xi \in \Xi,
\end{aligned}
\end{equation}
where $f_i$, $i=1,...,J$, are generic functions associated with each stage. 
For a two-stage stochastic program, $\mathcal{J}{=}\{2\}$ and $x^{2}$ maps directly to $y$. The optimization objective in a two-stage stochastic program generally is a function of both first- and second-stage variables. However, PH-based solution methods apply in situations where dependence of cost for one class of variables may be absent. 

The first term in (\ref{stoprog_obj}) represents the costs associated \emph{only} with the first stage, and thus $f_1$ is a function of the first-stage decisions $x^1$. The second term in (\ref{stoprog_obj}) represents the expected costs associated with subsequent stages $j{\in}\mathcal{J}$. For this term, $f_j$ is a function of all $j-1$ stage decisions $\vec{x}^{j-1}$ and sampled random realizations for all scenarios $s{\in}\mathcal{S}$. For a two-stage program, this is just a function of $x$ and $y_s$ for each scenario. A feasible solution requires all $j$-stage variables to be the same across all scenarios $s{\in}\mathcal{S}$, so we can only select first-stage variables $x$ that are feasible for \emph{all} random parameters considered. For a two-stage program, second-stage variables $y_s$ must be feasible for their associated scenario $s$ and the random variables realized for that scenario $\xi(s)$.

PH decomposes problem~\eqref{stoprog_obj} by scenario, in contrast to stage-based decomposition approaches such as the L-shaped method~\cite{vanslykewets69}. In a two-stage program, first-stage variables are decoupled, which effectively creates independent sub-problems corresponding to the individual scenarios; we denote the resulting scenario-specific first-stage variables as $x_s$. This yields an alternative version of an extensive form of a stochastic program, as follows:
\begin{subequations}
    \label{ef}
        \begin{equation}
            \label{ef_obj}
            \min_{x,y_s} 
            \sum_{s{\in}\mathcal{S}} \mathbb{P}(s) 
            \bigg{[} 
            f_1\big{(}x_s^1\big{)} + \sum\limits_{j=2}^{J}  f_j\Big{(}x_s^j;\vec{x_s}^{j-1},\vec{\xi}^j(s)\Big{)}
            \bigg{]}
        \end{equation}
        \begin{equation}
            \label{ef_nonant}
            x = x_{s},  \qquad \forall s \in \mathcal{S},\; x \in X_{\xi},\; \xi \in \Xi.
        \end{equation}
\end{subequations}

Recall that a valid solution to the overall problem is contingent on the first-stage variables agreeing across all scenarios $s{\in}S$. To ensure the optimization does not make decisions based on knowledge of the future, all scenario-specific first-stage variables must be linked together via non-anticipativity constraints to ensure $x = x_1 = ... = x_s$, as shown in \eqref{ef_nonant}. 

To solve this problem, PH iteratively reaches a solution that satisfies all non-anticipativity constraints and scenario-specific constraints. Let us refer to the iteration number as $v$. PH drops the non-anticipativity constraints \eqref{ef_nonant} and solves each scenario-based sub-problems independently. This step becomes trivially parallelizable since each sub-problem is separable from each other. The objective of each decomposed problem is appended with a proximal term, as shown in \eqref{prox}:
\begin{equation}
    \label{prox}
    \centering
    \frac{\rho}{2} \lvert\lvert x - \bar{x}^{(v)} \rvert \rvert^2.
\end{equation}
This term penalizes the objectives based on the magnitude of disagreement between all first stage variable values and also generally prohibits unbounded sub-problems. Here, $\rho$ is the quadratic penalty parameter. 

\begin{algorithm}[b]
    \caption{Progressive Hedging Algorithm for Optimizing Battery Sizing, Siting, and Operation}\label{alg:pha}
    \begin{algorithmic}[1]
        \State \textbf{Initialize:} $v \leftarrow 0$ and $w_s^{(j,v+1)} \leftarrow 0, \, \forall s \in \mathcal{S},j\in\mathcal{J}$
        \newline
        Compute $\forall s \in \mathcal{S}$
        \begin{center}
            $x^{(v+1)}_s \in \underset{x_s}{\text{argmin}} f_1\big{(}x^1_s\big{)} + \sum\limits_{j=2}^{J} f_t\Big{(}x^j_s;\vec{x}^{j-1}_s,\vec{\xi}^j(s)\Big{)}$ 
        \end{center}
        \State \textbf{Iteration Update:} $v \leftarrow v+1$
        \State \textbf{Aggregation:} $\bar{x}^{(v+1)} \leftarrow \underset{s{\in}\mathcal{S}}{\sum}\mathbb{P}(s)x^{(v)}_s$
        \State \textbf{Price Update:} $w_s^{(j,v+1)} \leftarrow w_s^{(j,v)} + \rho \Big{(} x_s^{(j,v)} - \bar{x}^{(j,v)} \Big{)} $
        \State \textbf{Decomposition:} 
        \newline
        Compute $\forall s \in \mathcal{S}$
        \begin{center}
            $x^{(v+1)}_s \in \underset{x_s}{\text{argmin}} f_1\big{(}x^1_s\big{)} + \sum\limits_{j=2}^{J} f_t\Big{(}x^j_s;\vec{x}^{j-1}_s,\vec{\xi}^j(s)\Big{)} $
        \end{center}
        \begin{center}
            $+ \sum\limits_{j=1}^{J-1}
            \big{[} w_s^{(j,v+1)\top}x^j_s + \frac{\rho}{2}||x^j_s - \bar{x}^{(j,v+1)}||^2
            \big{]} $
        \end{center}
        \State \textbf{Termination:} If criterion met, Stop. Otherwise $\rightarrow$ Step 2.
    \end{algorithmic}
\end{algorithm}

The general PH algorithm for a two-stage stochastic program is outlined in Algorithm \ref{alg:pha}. PH has convergence guarantees in the context of linear programs with continuous variables, and acts as a heuristic with gap-closing capabilities when discrete decision variables are involved. PH employs and updates scenario-specific weight vectors $w_s^{(j,v)}$ at each iteration $v$ of the algorithm. These vectors function similarly to dual weights, with the overall goal of pushing each sub-problem to optimality while \emph{simultaneously} encouraging agreement of first-stage variable values across all scenarios $s{\in}\mathcal{S}$. After solving the decomposed sub-problems, the aggregated first-stage variables $\bar{x}^{(j,v+1)}$ are calculated. These are used as a measure of consensus among all of the disaggregated first-stage variables $x_s$. Next, PH updates the price vector $w_s^{(j,v+1)}$ for all scenarios $s{\in}\mathcal{S}$. This weighting factor encourages first-stage variables that are far from the average of all scenario first-stage variables $\bar{x}$ to move closer to the average when solving the next iteration's sub-problems. There are certain properties of the price vectors that must be ensured to maintain algorithmic guarantees. PH maintains these requirements by the way price vectors are initialized and updated~\cite{rockafellar1991scenarios}. Sub-problems are solved again, and the iterative process continues until a specified convergence criteria is met. For example, this could be when the deviations between the dis-aggregated first-stage variable and their averages falls below a specified tolerance. 

\section{Application of Progressive Hedging to Battery Sizing, Siting, and Operating Problems}
\label{sec:methodology}

We use PH, as described in Section~\ref{sec:ph-algorithm}, to solve the battery sizing, siting, and operation optimization problem introduced in Section~\ref{sec:modeling}. We leverage the capabilities of the open-source Python package \texttt{mpi-sppy}~\cite{knueven2023parallel},
which is built on 1) the algebraic modeling language \mbox{Pyomo}~\cite{bynum2021pyomo} to specify optimization sub-problems and 2) MPI (Message Passing Interface) -- specifically, \texttt{mpi4py} -- to efficiently execute PH in parallel.

We adapt the scenario-based approach of PH to allow for decomposition by time periods $\mathcal{T}$, in addition to allowing for first-stage variables (relating to state-of-charge variables) that do not appear in every time period's sub-problem. We emphasize that this is a key modification relative to traditional progressive hedging algorithms to account for the temporal decomposition necessary to model the batteries' states-of-charge. We begin by decomposing the optimization model into a set of sub-problems $\mathcal{P}{=}\{1,...,P\}$, each representing a contiguous period of time from the full time horizon $\mathcal{T}$. In the WECC-240 case study described subsequently, we assign groups of contiguous days to form a total of $P=50$ time periods for roughly week-long subproblems. Each individual time $t$ within a time period $p$ represents a single hour. 

\begin{figure}[t]
    \centering
        \includegraphics[scale=0.30]{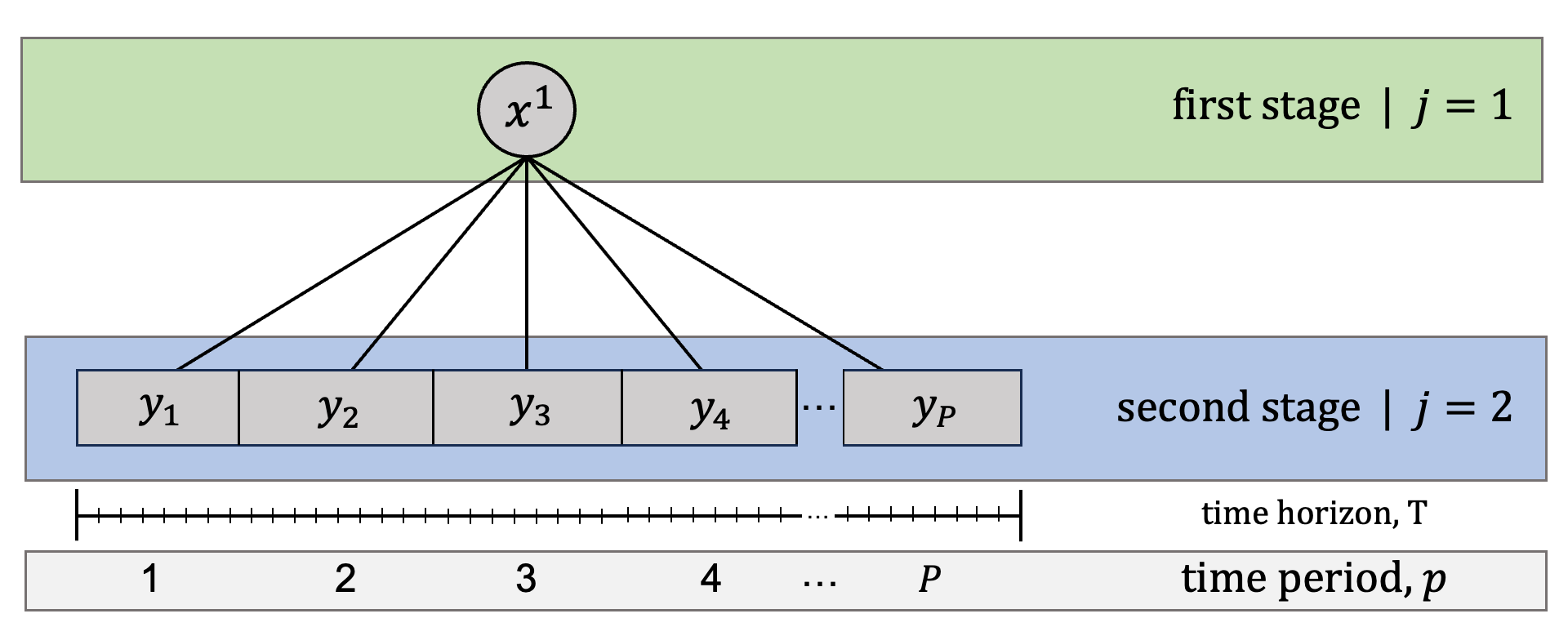}
        \caption{Time-period decomposition structure of battery sizing, siting, and operation model.}
        \label{fig:timeperiod-decomp}
        \vspace{-1em}
\end{figure}

The resulting decomposition structure is depicted in Figure~\ref{fig:timeperiod-decomp}, and is mathematically identical to a two-stage stochastic program decomposition by scenario. In the first stage, we have battery placement and sizing decisions, $x^n$; these must be non-anticipative relative to all time periods in $\mathcal{P}$. The first-stage cost $f_1$ is $C_{gen}$, per~\eqref{Cost-Opt}. All operational decisions (including load shedding, generator outputs, power flow, voltage angles, and states-of-charge) are second-stage variables that can vary by sub-problem. 

\begin{figure}[t]
    \centering
        \includegraphics[scale=0.34]{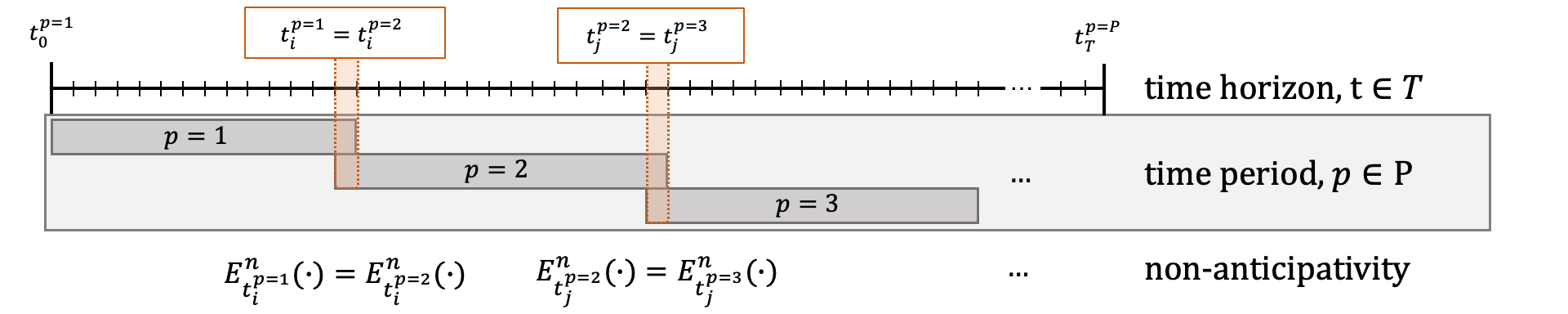}
        \caption{Battery state-of-charge decomposition structure in the battery sizing, siting, and operation model.}
        \label{fig:soc-breakdown}
\end{figure}

The decomposition discussed so far addresses all aspects of our optimization model with the exception of battery state-of-charge variables that couple time periods of adjacent sub-problems. We must ensure the states-of-charge of every placed battery are consistent at the start and end of temporally adjacent sub-problems. These additional complicating constraints are absent in traditional stochastic programming contexts. To address this final aspect of our optimization model in the context of PH-based decomposition, we extend the index set for all state-of-charge variables $E_t^n(\cdot)$ for each time period $p{\in}\mathcal{P}$ by including the last unit of time in the immediately preceding time period $p-1$. (No such modification is performed for the first time period.) This structure is depicted graphically in Figure \ref{fig:soc-breakdown}. These shared variables between temporally adjacent sub-problems are treated as additional first-stage variables. However, these variables do not appear in all sub-problems, requiring introduction of corresponding variable-specific probabilities (i.e., scenario weights in the context of the battery sizing, siting, and operation problem) -- specifically equal to 0.5 for sub-problems with the variable and 0 otherwise. Finally, we note that the state-of-charge variables are not referenced in the objective function (state-of-the-charge is not explicitly costed), such that determination of PH penalty parameters $\rho$ for such variables is more challenging~\cite{watson2011progressive}.

\begin{figure}[t]
    \centering
        \includegraphics[scale=0.38]{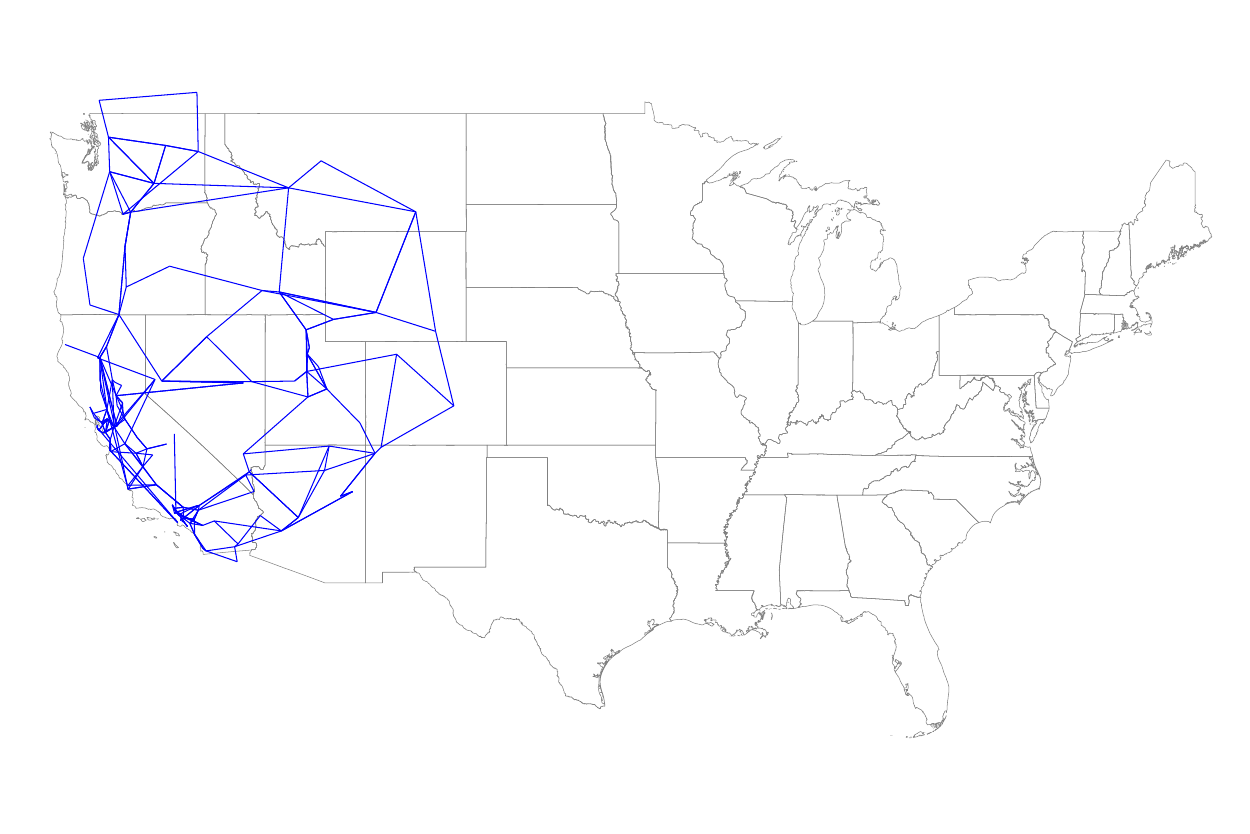}
        \caption{Geo-location of the WECC-240 system network.}
        \label{fig:network locations}
\end{figure}

\begin{figure}[h]
\centering
\begin{subfigure}{.45\textwidth}
  \centering
    \includegraphics[width=.8\linewidth]{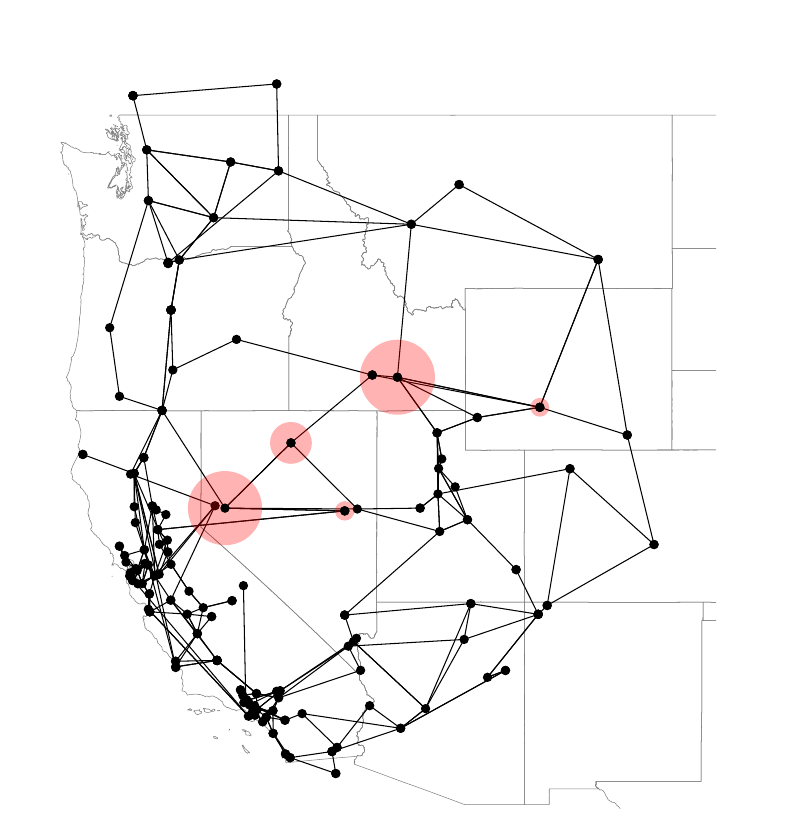}  
  \caption{April 2021}
\end{subfigure}
\begin{subfigure}{.45\textwidth}
  \centering
  \includegraphics[width=.8\linewidth]{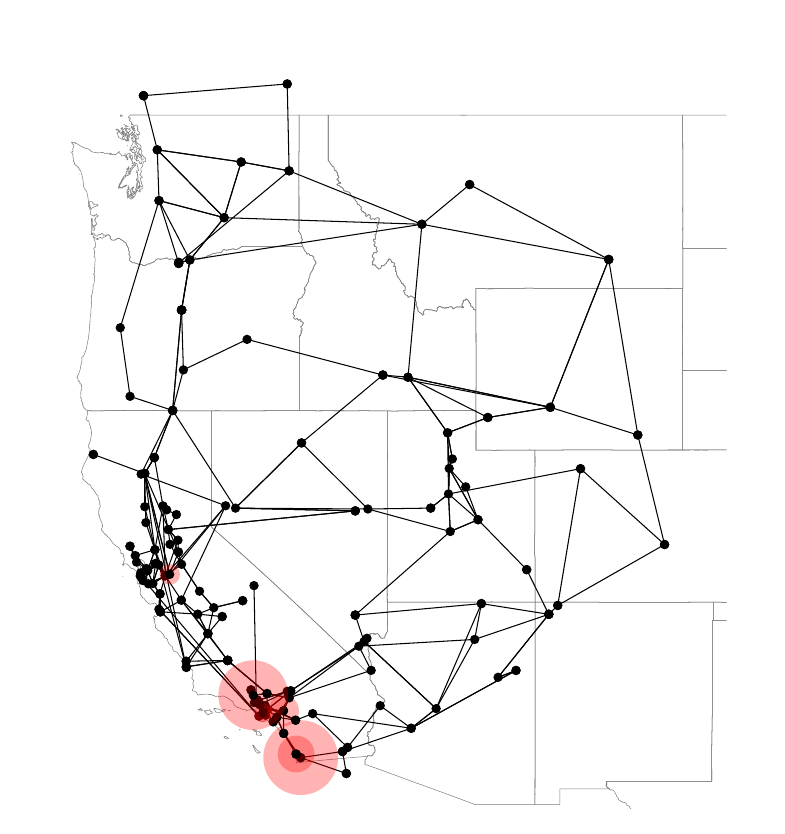}  
  \caption{June 2021}
\end{subfigure}
\caption{Optimal battery placements for the WECC network in April 2021 (top) and in June 2021 (bottom). Red circles are sized proportionally to the number of batteries placed at that bus with the largest circle representing 4 batteries.}
\label{figure:deterministic battery comp}
\end{figure}

\section{Case Studies}\label{sec:results}

To demonstrate the capabilities of our algorithm, we consider a synthetic test network geolocated in the western United States with 240 buses (WECC-240)~\cite{WECC_case}, for which we associate temporally and spatially varying wildfire risk data from the USGS Wind-Enhanced Fire Potential Index~\cite{USGS_WFPI}. Wildfire data is defined for the end of April through November. The network topology and electrical parameters are adopted from~\cite{pglib-opf} augmented with data originating from~\cite{barrows2019ieee} and ~\cite{price2011reduced}. To avoid infeasibility, power inputs represented as DC lines modeled as negative loads were deactivated for these simulations and slack generation is included at each bus. The geo-located structure of the WECC-240 network topology is provided in Figure~\ref{fig:network locations}. 
We next consider both direct solution of the optimization model described in Section~\ref{sec:modeling} and solution using PH-based decomposition (per Sections~\ref{sec:ph-algorithm} and~\ref{sec:methodology}). 

\subsection{Solution of the Extensive Form}\label{sec:determ results}

We first conduct tests on the WECC-240 case using the full extensive form model outlined in Section~\ref{sec:modeling}, i.e., considering direct solution of~\eqref{stoprog_obj} without decomposition as in Algorithm~\ref{alg:pha}. The corresponding models were solved using Gurobi~10.0.0~\cite{gurobi}. To implement the optimization formulations, we used Julia 1.8.0~\cite{bezanson2017julia} with JuMP v1.11.1~\cite{DunningHuchetteLubin2017} along with the data input functionality of PowerModels.jl v0.19.8~\cite{coffrin2018}. Computations were conducted on the Partnership for an Advanced Computing Environment (PACE) at the Georgia Institute of Technology~\cite{PACE}. Each case was run on one node with 8 cores with 24 GB of memory. Each node has Dual Intel Xeon Gold 6226 CPUs @ 2.7 GHz. 

Figure~\ref{figure:deterministic battery comp} shows the resulting battery locations during nominal operation (for a month in April of 2021) and during a period of high wildfire risk (for a month in June of 2021). From these tests, we observe that optimal placement location varies based on the time of year. 
Given the computational difficulties with long time horizons, the deterministic model cannot be used to find optimal placement values over the span of a full year. For the one-month extensive form in June 2021, the model contains over 1.2 million constraints, 1.9 million variables, and 5 million nonzero values. The one-month simulations took nearly 25 minutes of wall-clock time. A one-year simulation was terminated without a solution after a 24-hour build time. 
Instead, we turn to the PH algorithm described in Sections~\ref{sec:ph-algorithm} and~\ref{sec:methodology}.  Results from the temporally decomposed problem are discussed next. 

\begin{figure}[t]
    \centering
        \includegraphics[width=0.75\linewidth]{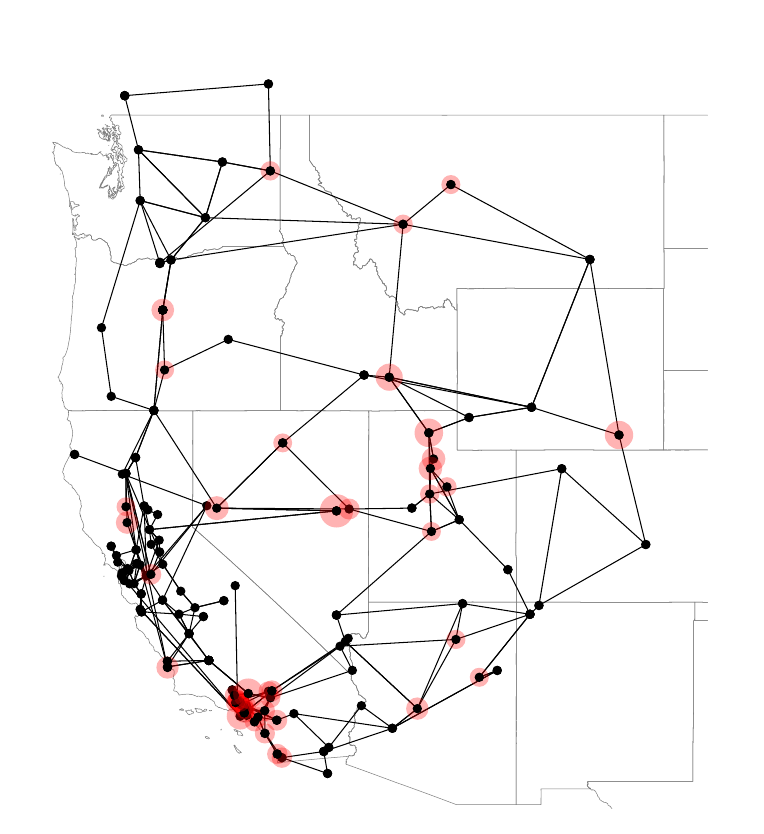}
        \caption{Placement decisions on the WECC-240 network from the full-year simulation decomposed into 120 time periods. Red circles are sized proportionally to the number of batteries placed at that bus with the largest circle representing 1.05 batteries.}
        \label{fig:wecc full year locs}
\end{figure}

\subsection{Decomposition Results} \label{sec:decomp results} 

For PH, we decompose the full one-year time horizon into 120 time periods, most with 72 hours, and a handful of scenarios including an additional 24 hours, to model the full year. We report PH solutions that are fully non-anticipative, i.e., are valid incumbents for the extensive form optimization model. Following the methodology outlined in Section~\ref{sec:methodology}, we solved a decomposed, one-year horizon battery sizing, siting, and operating problem for the WECC-240 network. We used Gurobi v10.0.2 \cite{gurobi} to solve this problem in Python v3.9.12 and modeled the optimization formulation in Pyomo v6.6.1~\cite{bynum2021pyomo}. We employed \texttt{mpi-sppy} v0.12~\cite{knueven2023parallel} to execute PH in parallel, as outlined in Section~\ref{sec:ph-algorithm}. This was executed on the quartz HPC at Lawrence Livermore National Laboratory. We executed this problem using 60 nodes of the HPC. Each node has two 18-core Xeon E5-2695 processors (2.1 GHz) and 128 GB of memory. Furthermore, we decomposed the problem into 120 scenarios. We used 240 MPI ranks to solve the problem, with 120 dedicated to running the PH algorithm on each sub-problem and 120 dedicated to executing an asynchronous incumbent-finder. (A rank is a unique number that identifies each parallel process.) Each Gurobi solve was limited to 36 threads, and the $\rho$ parameter was set to a constant value of 0.001. 

For the decomposed problem, the reported solution was found after 50 iterations of the PH algorithm and incumbent finders. We reached a relative optimality gap of 0.023\% and an absolute gap of 0.367 in under 70 minutes. Note, a total slack generation of 0.0106 p.u. was included in the solution throughout the year across all buses to ensure feasibility. The extensive form of a one-year run, as mentioned, did not \emph{build} within 24 hours. 
 
\begin{figure}[t]
    \centering
        \includegraphics[scale=0.38]{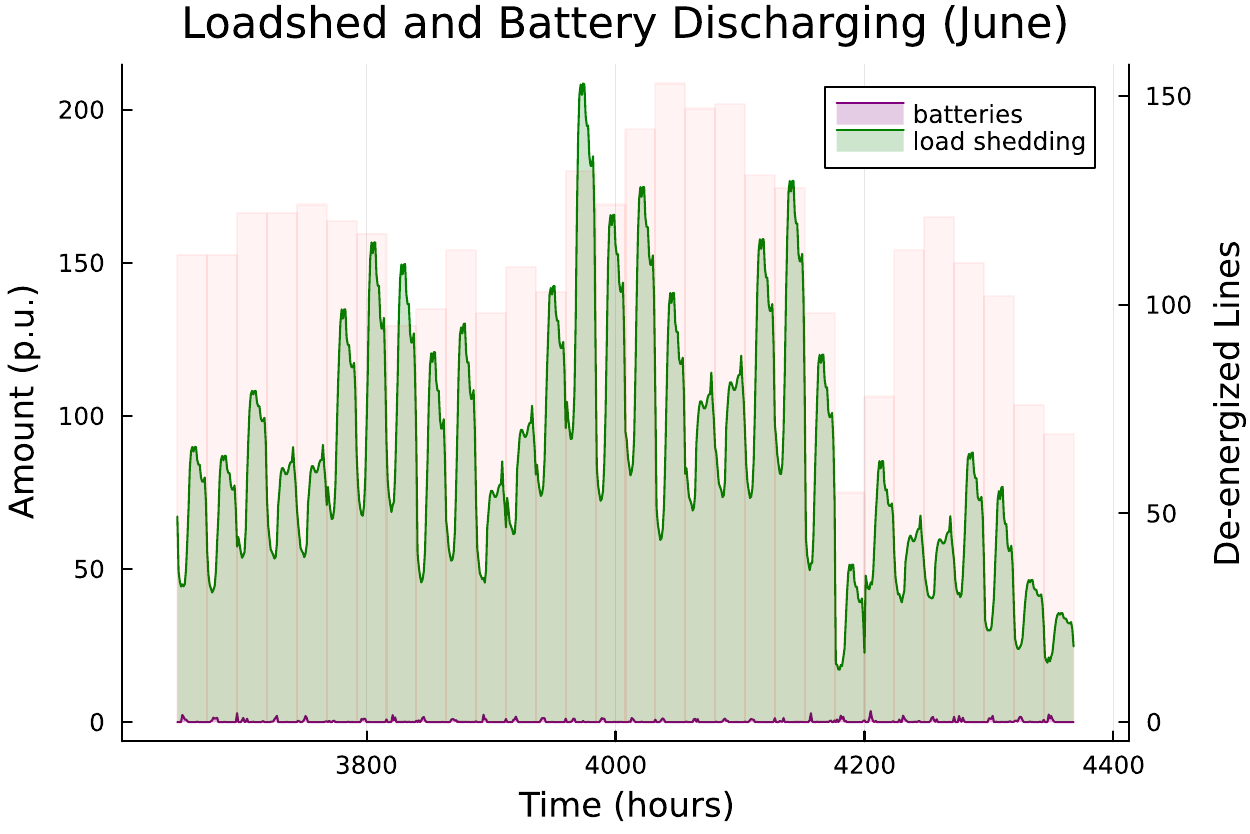}
        \caption{Load shedding and battery discharge for the WECC-240 network in June of 2021. Note that battery discharging decisions and load shed are both used to serve power in addition to generation (not shown).}
        \label{fig:june power sources}
        \vspace{-1em}
\end{figure}

\begin{figure*}[ht]
\centering
\begin{subfigure}{.48\textwidth}
  \centering
    \includegraphics[width=\linewidth]{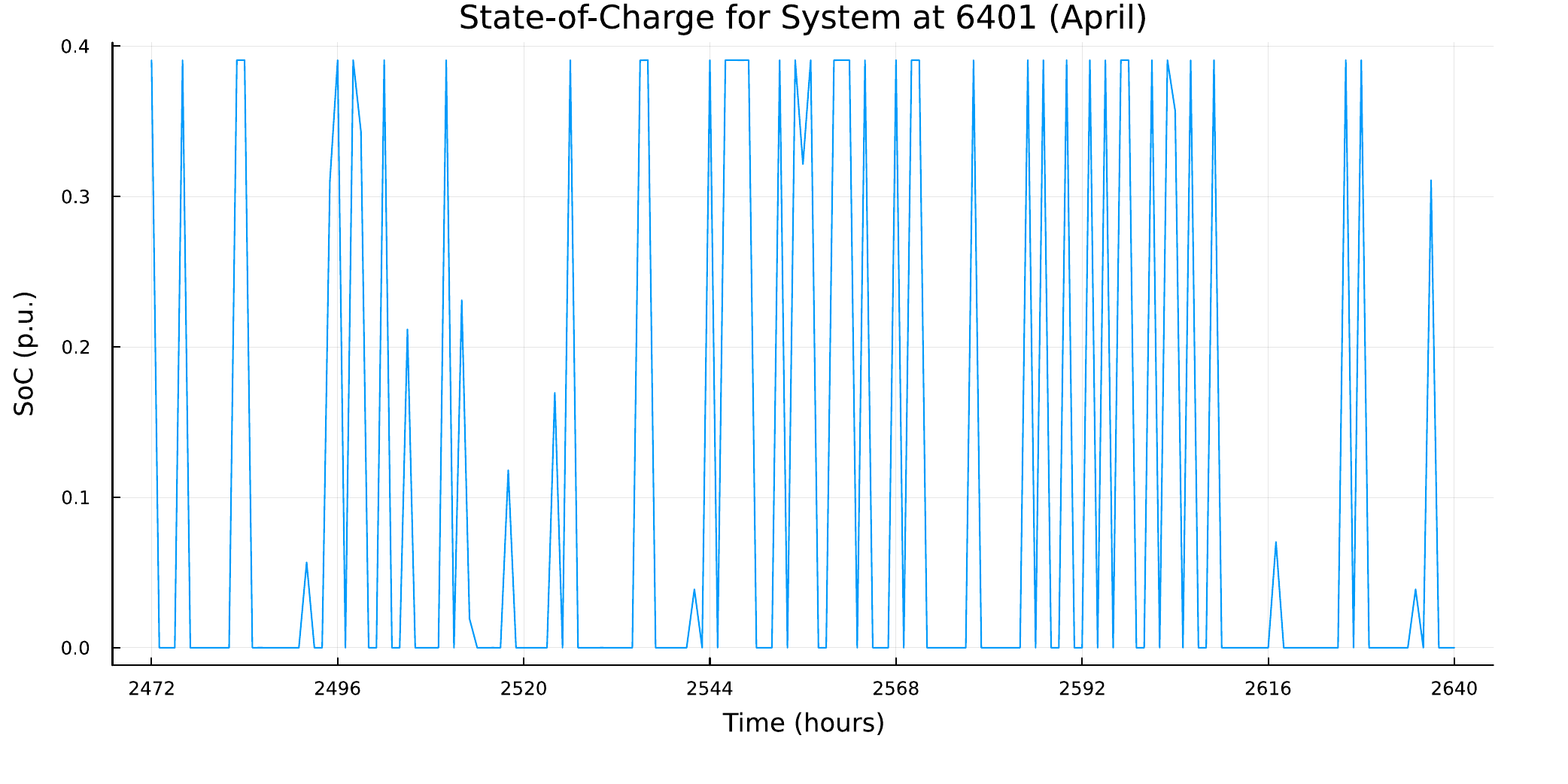}  
\end{subfigure}
\begin{subfigure}{.48\textwidth}
  \centering
  \includegraphics[width=\linewidth]{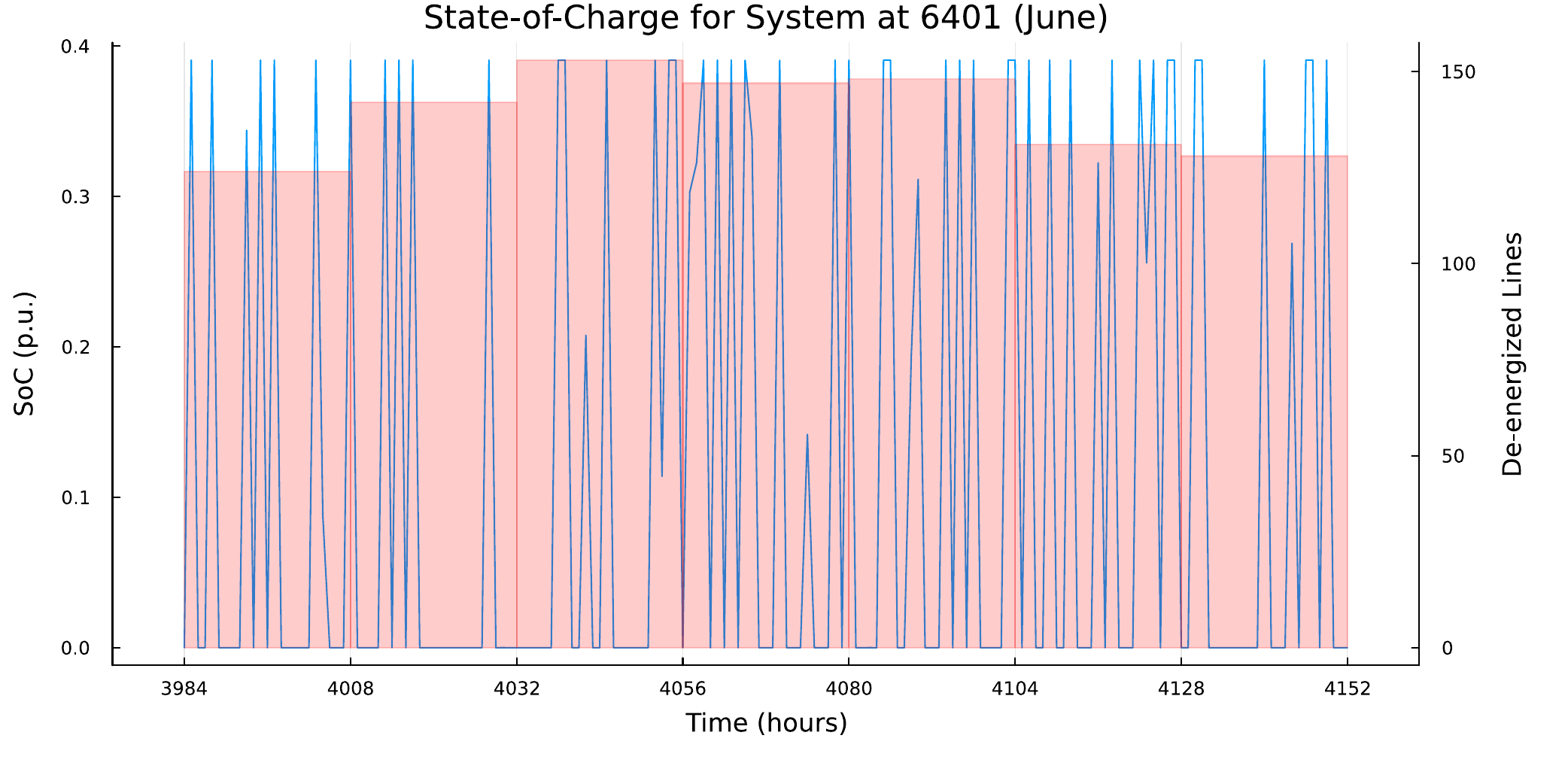}  
\end{subfigure}
\caption{State-of-Charge of batteries at bus 6401 on the WECC-240 network in a week of April 2021 (left) and a week of June 2021 (right) with red bars showing the number of de-energized transmission lines on each day.}
\label{figure:distributed soc comp}
\end{figure*}

Comparing the placement locations from the full-year decomposition, seen in Figure~\ref{fig:wecc full year locs}, to the the month-based placements in Figure~\ref{figure:deterministic battery comp}, we observe that the final sites have some overlap with decisions from the extensive model of April \emph{and} the placement decisions in June. Results from April and June are shown as a subset of the results of the full year. As can be seen in Figure \ref{fig:june power sources}, during the month of June, there is considerable load shedding during days with many lines de-energized. Battery operation at bus 6401 in the WECC-240 can be seen in Figure~\ref{figure:distributed soc comp}. It is interesting to note that there is very minimal difference in the operation of the batteries between April and June in these results. This might be explained by the selected battery shown largely responding to local demand and not being significantly influenced by line de-energizations. As seen in Figure~\ref{figure:deterministic battery comp}, batteries are most helpful in the southwestern US for the June 2021 wildfire risk profile. When looking at the battery placements based only on April and June, we see drastically different results. The results from the full year decomposed run reflects this as battery installations are distributed throughout the network to aid in operations at all times of the year in different regions. %

\section{Conclusion}
\label{sec:conclusion}

We have introduced a new approach to solving utility-scale battery sizing, siting, and operations optimization problems, considering the mitigation of wildfire risk and associated load shedding. Direct approaches to solving an extensive form of the optimization model fail to scale to significant time horizons, e.g., on the order of a year. Long time horizons are critical to capture operations in both nominal and wildfire seasons. We introduce a decomposition approach based on a progressive hedging (PH) algorithm. Our approach allows us to capture both a standard planning/operations decomposition structure, but also non-anticipative variables (state-of-charge in our model) that couple sub-problems representing temporally adjacent operational periods. Implemented in a scalable open-source stochastic programming library (\texttt{mpi-sppy}) and executed using parallel computing clusters, our PH-based decomposition yields near-optimal battery placements across a year-long time horizon in tractable run times. The ability to consider long time horizons is shown to be critical, in that the resulting battery placements can differ from placements obtained using shorter time horizons. Our implementation of PH with temporal decomposition allows for benefits in solve time as well as considerable benefits in parallelizing the build time for large models. 

In addition, these tools can be used for other forms of infrastructure upgrades such as line undergrounding. Similarly, longer time scales could be considered. Instead of making investment decisions based on one year, a rolling-budget schedule could be introduced that allows infrastructure to be placed over multiple years with changing load demand and varying wildfire risk. Given the nature of the progressive hedging algorithm, this formulation could also be extended to consider multiple wildfire risk profiles and other uncertainties within each time period. 

\section{Acknowledgements}
This document was prepared as an account of work sponsored by an agency of the United States government. Neither the United States government nor Lawrence Livermore National Security, LLC, nor any of their employees makes any warranty, expressed or implied, or assumes any legal liability or responsibility for the accuracy, completeness, or usefulness of any information, apparatus, product, or process disclosed, or represents that its use would not infringe privately owned rights. Reference herein to any specific commercial product, process, or service by trade name, trademark, manufacturer, or otherwise does not necessarily constitute or imply its endorsement, recommendation, or favoring by the United States government or Lawrence Livermore National Security, LLC. The views and opinions of authors expressed herein do not necessarily state or reflect those of the United States government or Lawrence Livermore National Security, LLC, and shall not be used for advertising or product endorsement purposes. This work was performed in part under the auspices of the U.S. Department of Energy by Lawrence Livermore National Laboratory under Contract DE-AC52-07NA27344 and was supported by both the LLNL LDRD Program under Project 22-SI-008 and the US Department of Energy Office of Electricity's Advanced Grid Modeling program. D.K. Molzahn and R. Pianksy acknowledge support from the NSF AI Institute for Advances in Optimization (AI4OPT), \#2112533. The authors would also like to thank Gurobi for providing the academic license used to run tests and D.L. Woodruff at University California -- Davis with assistance on implementation. G. Stinchfield and R. Piansky are co-first authors.


\bibliographystyle{IEEEtran}
\bibliography{IEEEabrv,refs.bib}

\end{document}